\def\BEQ{\begin{eqnarray}} 
\def\EEQ{\end{eqnarray}} 
\def\BML{\begin{mathletters}} 
\def\EML{\end{mathletters}} 
\def\BE{\begin{equation}} 
\def\EE{\end{equation}} 
\def\NN{\nonumber} 
\let\SCR=\scriptstyle 
 
\def\d#1{{\partial \over \partial #1}} 
\def\bra#1{\left\langle #1} 
 
\def\ket#1{#1 \right\rangle}

\def\moy#1{{\left\langle #1 \right\rangle}} 
\def\vec#1{{\bf #1}} 
 
\def\be{{\beta}} 
\def\ga{{\gamma}} 
\def\Ga{{\Gamma}}

\def\si{{\sigma}} 
\def\om{{\omega}} 
 
\def\up{{\uparrow}} 
\def\dow{{\downarrow}} 
\def\w{{i\om _n}} 
 
\documentstyle[twocolumn,prl,aps,floats]{revtex}
\input epsf.tex
\begin{document} 
\twocolumn[\hsize\textwidth\columnwidth\hsize\csname@twocolumnfalse%
\endcsname
\title{Strong-Coupling Expansion for the Hubbard Model} 
\author{St\'ephane Pairault\cite{Stef} , David S\'en\'echal, and 
 A.-M. S. Tremblay} 
\address{D\'epartement de 
Physique and Centre de Recherche en Physique du Solide} 
\address{Universit\'e de Sherbrooke, Sherbrooke, Qu\'ebec, Canada J1K 2R1.} 
\address{\rm preprint CRPS 97-25, published in Phys. Rev. Lett. {\bf 80},
5389  (1998).} 
\maketitle 
\begin{abstract}%
A strong-coupling expansion for models of correlated electrons in any
dimension is presented. The method is applied to the Hubbard model in $d$
dimensions and compared with numerical results in $d=1$. Third order expansion
of the Green function suffices to exhibit both the Mott metal-insulator
transition and a low-temperature regime where antiferromagnetic correlations
are strong. It is predicted that some of the weak photoemission signals
observed in one-dimensional systems such as $\rm SrCuO_2$ should become
stronger as temperature increases away from the spin-charge separated state.%
\end{abstract}

]

Organic conductors, cuprate ladder compounds and High-$T_c$
superconductors are but a few of the condensed matter systems currently
driving the intense experimental and theoretical efforts on strongly
correlated electrons in low dimension ($d=1$ or $d=2$). Angle-resolved
photoemission experiments (ARPES) on 2D and 1D materials
\cite{Wells95,Kim96} are beginning to probe the spectral weight $A(\vec
k,\om)$, but have not yet given definitive answers to questions of
prior interest such as spin-charge separation or the relation between
the Mott transition and antiferromagnetic (AF) correlations. On the
theoretical side, the Hubbard model~(HM)~\cite{Hubbard63} is the
simplest one that includes the interplay between the strong screened
Coulomb repulsion and the kinetic band energy. In one dimension, the
exact solution of the HM\cite{Lieb68} does not allow actual calculation
of correlation functions, but several numerical studies
\cite{Preuss94,Preuss95,Bulut94,Favand97} have investigated the
one-particle spectral function that is observed in ARPES. Other methods
(bosonization\cite{Emery79,Voit94,Voit96}, renormalization
group\cite{Solyom79,Bourbon91}, conformal field theory\cite{Frahm90})
have led to important nonperturbative results, but in addition to being
restricted to the lowest energy excitations, they involve parameters
that are absent from the microscopic Hamiltonian. For the
$d=\infty$ \cite{Kotliar92,Georges93,Pruschke93,Georges96}
case, all quantities of interest can be calculated in an essentially
exact way, but the extrapolation to low dimension is problematic.
Several attempts to develop systematic strong-coupling
expansions\cite{Bartkowiak92,Metzner91} did not yield the spectral
weight.

\begin{figure}
\epsfxsize 8cm\centerline{\epsfbox{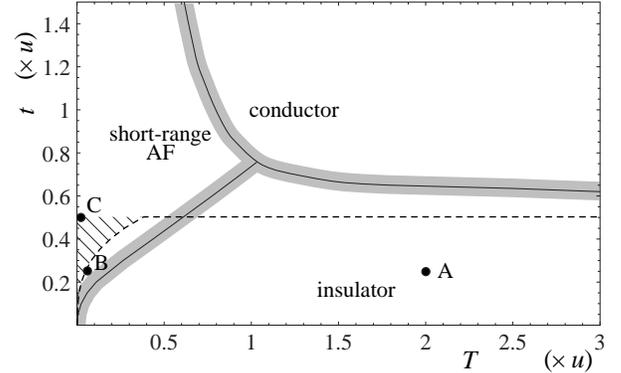}}
\caption{
Crossover diagram of the half-filled 1D Hubbard model
with Coulomb repulsion $U=2u$ and hopping $t$.}
\label{phases}
\end{figure}

The purpose of this letter is two-fold. First, we construct a
strong-coupling perturbation theory that can be applied to a number of
models in {\it any} dimension, and, second, we use it to compute the
Green function of the half-filled HM. This allows us to discuss the
Mott transition from the viewpoint of the density of states. The
effects of antiferromagnetic correlations on $A(\vec k,\om)$ are
discussed, for simplicity, only in 1D. Despite the absence of phase
transitions in 1D, the qualitative behaviour of
$A(\vec k,\om)$ allows us to define crossovers between regions of
parameter space where the system behaves somewhat like a metal, an
insulator or a short-range antiferromagnet. The results are summarized
by the crossover diagram of Fig.~\ref{phases}, which shows some
analogies with one of the published $d=\infty$ phase diagrams
\cite{Pruschke93}. We conclude with a prediction for 1D systems of
current experimental interest~\cite{Kim96}.

First, let us present the strong-coupling expansion itself. Consider a
Hamiltonian ${\cal H}={\cal H}^0+{\cal H}^1$, where the unperturbed part
${\cal H}^0$ is diagonal in a certain variable $i$ (say a site variable), and
let us denote collectively by $\si$ (say a spin variable) all the other
variables of the problem. This Hamiltonian involves fermions, and is supposed
to be normal ordered in terms of the annihilation and creation operators
$c_{i\si}^{(\dag)}$. ${\cal H}^0$ may be written as a sum over $i$ of on-site
Hamiltonians involving only the operators $c_{i\si}^{(\dag)} $ at site $i$:
${\cal H}^0=\sum_i h_i (c_{i\si}^{\dag} ,c_{i\si})\>.$ For a strong-coupling
expansion of the HM, ${\cal H}^0$ is the atomic limit, namely
$h_i (c_{i\si}^{\dag} ,c_{i\si})=
Uc_{i\up}^{\dag}c_{i\dow}^{\dag}c_{i\dow}c_{i\up}\>$ (we will often use
$u=U/2$ for convenience). We suppose that the perturbation ${\cal H}^1$ is a
one-body operator of the form
${\cal H}^1 =\sum_\si \sum_{ij} V_{ij} c_{i\si}^{\dag}c_{j\si} \>$. For the
HM, ${\cal H}^1$ is the kinetic term. Introducing the Grassmann field
$\ga_{i\si}(\tau), \ga_{i\si}^{\star}(\tau)$, the partition function at some
temperature $T=1/\be$ may be written in the Feynman path-integral formalism:
\BEQ
\label{partition1}
Z&=&\int [d\ga^{\star} d\ga ] \exp -\int_0^{\be} d\tau \Bigg\{
\sum_{i\si}\ga_{i\si}^{\star}(\tau)\left( \d\tau -\mu \right)
\ga_{i\si}(\tau) \NN \\  && +
\sum_i h_i (\ga_{i\si}^{\star}(\tau) ,\ga_{i\si}(\tau))
+ \sum_{ij\si} V_{ij} \ga_{i\si}^{\star}(\tau)
\ga_{j\si}(\tau) \Bigg\} \> .
\EEQ
We use
the letters ($a,b,...$) to denote sets such
as $(i,\si ,\tau)$,
for instance :
\BEQ
\NN
\int_0^{\be}d\tau \sum_{ij\si} V_{ij} \ga_{i\si}^{\star}(\tau)
\ga_{j\si}(\tau)
=\sum_{ab} V_{ab} \ga_a^{\star}\ga_b \>.
\EEQ

A first difficulty arises: There is no Wick theorem because
${\cal H}^0$ is quartic instead of quadratic.
We solve this problem by means of
a Grassmannian Hubbard-Stratonovich transformation,~\onlinecite{Boies95}
which consists in expressing the perturbation part of the action
in Eq.~(\ref{partition1})
as a Gaussian integral over an auxiliary
Grassmann field $\psi_{i\si}(\tau), \psi_{i\si}^{\star}(\tau)$.
Then, the integral over the original variables can be performed and
$Z$ can be rewritten in the form:
\BEQ
\NN
Z=\int [d\psi^{\star}d\psi] \exp -\left\{ S_0[\psi^{\star},\psi ]
+\sum_{R=1}^{\infty}S_{\rm int}^R[\psi^{\star},\psi ] \right\} \>.
\EEQ
The action has a free part
\BEQ
\NN
S_0[\psi^{\star},\psi ]=-\sum_{ab}\psi^\star_a \left( V^{-1} \right)_{ab} \psi_b \>,
\EEQ
and an infinite number of interaction terms
\BEQ
\NN
S_{\rm int}^R[\psi^{\star},\psi ]={-1\over (R!)^2}\sum_{\{ a_l,b_l\} }{}'
\psi_{a_1}^\star .. \psi_{a_R}^{\star}\psi_{b_R}..\psi_{b_1}
G^{R\rm c}_{\SCR a_1..a_R\atop\SCR b_1..b_R},
\EEQ
where the $G^{R\rm c}_{\SCR
a_1..a_R\atop\SCR b_1..b_R}=\moy{\ga_{a_1}..\ga_{a_R}
\ga_{b_R}^\star ..\ga_{b_1}^\star}_{0,{\rm c}}$ are the connected correlation
functions of the unperturbed system. The primed summation reminds us that the
fields in each term share the same value of the site index. We may now use
Wick's theorem and usual perturbation theory for the $\psi$'s, the free
propagator being $V$, and the vertices being the $G^{R\rm c}$'s. The number of
auxiliary field propagators determines the order in $V$ ($\vert V_{ij}\vert
=t$ for the HM) of a given diagram. Finally, the relation between the Green
function
${\cal G}_{ab}=-\moy{\ga_a\ga_b^{\star}}$ of the original fermions and that of
the auxiliary field
${\cal V}_{ab}=-\moy{\psi_a\psi_b^{\star}}$, is (in matrix form)
$
{\cal G}=-V^{-1}+V^{-1}{\cal V}V^{-1}\>.
$
If $\Ga$ denotes the self-energy of the $\psi$'s, one has
$
{\cal G}=\left( \Ga^{-1}-V \right)^{-1}\>.
$


The above method was applied to the HM
\BEQ
\NN
{\cal H}=
2u\sum_i c_{i\up}^{\dag}c_{i\dow}^{\dag}c_{i\dow}c_{i\up}
-t\sum_{\bra i , \ket j \si} ( c_{i\si}^{\dag}c_{j\si} +{\rm H.c.})
\EEQ
at half-filling up to order $t^3$.
The result for ${\cal G}$ is a rational function of $\w$:
\BEQ
\label{Green1}
{1\over{\cal G}(\vec k ,\w)}= 2t c(\vec k)
+ \Bigg\{ {\w\over(\w)^2-u^2}+
{6dt^2u^2\w\over\left( (\w)^2-u^2 \right)^3} \NN \\
+6t^3c(\vec k)\left( {{\be u\over 2}\tanh{\left({\be u\over 2}\right) }\over
\left((\w)^2-u^2\right)^2}+{u^2\left(2(\w)^2-u^2\right)\over
\left((\w)^2-u^2\right)^4}
\right)
\Bigg\}^{-1},
\EEQ
where $d$ is the dimension of the hypercubic lattice, and
$c(\vec k)=\sum_{m=1}^d \cos (k_m)$. Here we face a second difficulty, namely
that ${\cal G}(\vec k,\w)$ has pairs of complex conjugate poles. This violates
the Kramers-Kr\"onig relations and leads to negative spectral weight. Note
that even in weak-coupling theory, truncation of the series for ${\cal G}$
leads to high-order poles giving negative spectral weight. Since we only know
${\cal G}$ up to order $t^3$, any function having the same Taylor expansion as
${\cal G}$ to this order is {\it a priori} as good an approximation. A
physically acceptable solution should be causal and have a positive spectral
weight, that is, be a sum of simple real poles with positive residues. We call
such a function Lehmann representable (LR).

In order to obtain a LR approximation, we need the following theorem, reported
in Ref.~\onlinecite{Gilewicz78}: A {\it rational} function is LR {\it if and
only if} it can be written as a finite Jacobi continued fraction
\BEQ
\label{Jacob1}
{\cal G}_J (\w)={a_0\over \w +b_1 -}\>
{a_1\over\w +b_2-}\>...\>
{a_{L-1}\over \w +b_L}\>,
\NN
\EEQ
with $b_l$ real and $a_l>0$ (thereafter conditions CO).

According to this theorem, the exact Green function of any finite system is a
Jacobi continued fraction, whose coefficients, functions of the hopping $t$,
verify conditions CO. If we expand the exact ${\cal G}_J $ in powers of $t$ to
some finite order, which is what a strong-coupling expansion does, we destroy
its continued fraction structure. If instead we replace
$a_l(t)$ and $b_l(t)$ in ${\cal G}_J $ by their expansion to some finite
order, the result should be LR since we expect conditions CO to hold for the
truncated coefficients (at least for $t/u$ small).

Therefore, to obtain a LR approximation, we seek frequency-independent
$a_l(t)$ and $b_l(t)$, such that ${\cal G}_J $ and $\cal G$ have the same
expansion up to order $t^3$. Equating the series in $t$ for $\cal G$ and for
${\cal G}_J$ at all frequencies determines uniquely the leading terms in the
$t$ expansion of $a_l(t)$ and $b_l(t)$. As soon as some
$a_l(t)$ is found to be zero up to the required precision necessary to obtain
the $t^3$ term of ${\cal G}_J$, all $a_p(t)$ and $b_p(t)$, $p>l$ become
unecessary.

The above procedure generalizes what is done in weak-coupling theory. There,
Wick's theorem allows a resummation of one-particle reducible diagrams, which
gives Dyson's equation. If the self-energy is LR, ({\it i.e.}, has an
underlying continued fraction structure), the Green function inherits this
property due to the form of the weak-coupling free propagator.


We were able to deduce from Eq.~(\ref{Green1}) the following
continued fraction
\BEQ
\label{Jacob2}
{\cal G}_J (\w)&=&{1\over
\w +2 t c(\vec k) -}\>{u^2\over
\w-{3\be t^3}\tanh{\left({\be u\over 2}\right) }c(\vec k)/u -} \NN\\
&& {6 dt^2\over
\w -2 t c(\vec k)/d -}\>{u^2\over
\w +t c(\vec k)/d}
\>,
\EEQ
which verifies the conditions CO, and has exactly the same Taylor expansion as
$\cal G$ up to order $t^3$ included. This means that {\it all} the
moments\cite{Moments} of ${\cal G}_J $ are the same as those of the exact
solution except for terms of order~$t^4$. Furthermore, any LR rational
function sharing this property reduces to a continued fraction whose
coefficients differ from those of Eq.~(\ref{Jacob2}) only by terms smaller
than the precision achieved here~\cite{precision}.

Expansion to order $t^3$ for the half-filled HM suffices to exhibit both the
Mott transition and the effect of AF correlations on the spectral weight
$A(\vec k,\om)=\lim_{\>\eta \rightarrow 0^+}-2\>{\rm Im}\>{\cal G}(\vec k,\om
+i\eta)$. There is no rigorous definition of the Mott transition in terms of
one-particle properties, but one can use, as a heuristic criterion, the
appearance of spectral weight at zero frequency. In the density of states
$N(\om)=\int_{-\pi}^{\pi}A(\vec k,\om) d^d\vec k/(2\pi)^d$, as $t$ increases
from zero, the two symmetric Hubbard bands located at $u$ and $-u$ in the
atomic limit widen, and eventually mix for $t$ beyond some critical value. The
latter may be obtained by demanding that a pole of $\cal G$ crosses the Fermi
level for some $\vec k$. For $T\rightarrow\infty$, the critical
value of $t$ is
$t_{\rm c}=u{\sqrt{1+\sqrt{1+12d^2}}/(2d\sqrt{3}})$~\cite{d-infty}.
This gives
$U_{\rm c}\simeq 3.2 t$ for $d=1$, to be compared with
$U_{\rm c}\simeq 3.5t$ found in the Hubbard-III\cite{Hubbard63} approximation.
At finite $T$, we cannot calculate $t_{\rm c}$ analytically, but
Fig.~\ref{phases} sketches a numerical evaluation (for $d=1$) in the ($T,t$)
plane of the line where the gap vanishes. The value of $t_{\rm c}$ grows upon
lowering $T$, and there is no Mott transition at zero temperature, in
agreement with the exact result of Ref.~\cite{Lieb68}.

\begin{figure}
\epsfxsize 8cm\centerline{\epsfbox{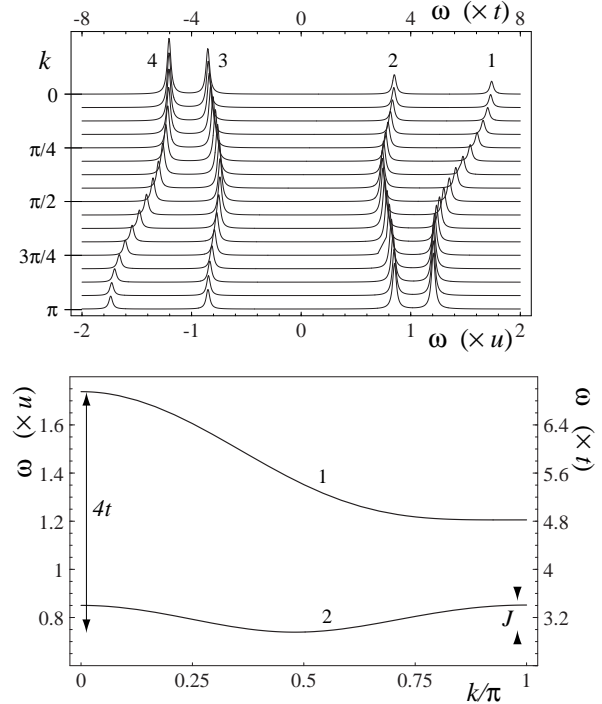}}
\caption{
(above)~Spectral function $A(k,\om)$ for point B
($t=0.25u,\>T=0.06u$ or $U=8t,\>T=0.24t$).
$\eta$ (see text) was set to $0.02$.
(below)~Dispersion relation of the peaks.
}
\label{spec1}
\end{figure}

The effects of AF correlations show up at low T, as illustrated in
Fig.~\ref{spec1} by the plot of $A(k,\omega)$ for point B of Fig.~\ref{phases}
($\bf k$ becomes $k$ because we discuss the 1D case for definiteness).
$A(k,\om)$ has four delta peaks (a finite width $\eta $ is added for clarity)
given by dispersion relations $\om_i(k)$, ($i$=1 to 4 as in Fig.~\ref{spec1}).
The spectral weight is an even function of $k$, and particle-hole symmetry
ensures that $A(k+\pi,-\om)=A(k,\om)$. While at small $t$ and high $T$ (point
A), $\omega_2(k)$ was minimum for $k=0$, when $T$ is lowered down to point B,
the minimum of $\omega_2(k)$ moves continuously from $k=0$ towards
$k=\pi/2$ (Fig.~\ref{spec1}), and peak 2 loses weight for values of $k$ much
smaller than $\pi/2$. These changes reflect the AF short-range order
that gradually builds up when $T$ becomes smaller than the AF superexchange
$J=2t^2/u$ of the equivalent $t-J$ model. The approximate cell doubling in
direct space translates into a nearly $\pi$-periodic dispersion for peak 2,
although the $2\pi$-periodicity of its weight and of $\om_1(k)$ reminds us
that the state remains paramagnetic. This is why we chose to define the AF
crossover line of Fig.~\ref{phases} as the points where $k=0$ ceases to be the
minimum of $\om_2(k)$. In this regime, the width of band 2 is of order
$J=2t^2/u$ whatever the value of $t$, supporting the above interpretation.

\begin{figure}
\epsfxsize 8cm\centerline{\epsfbox{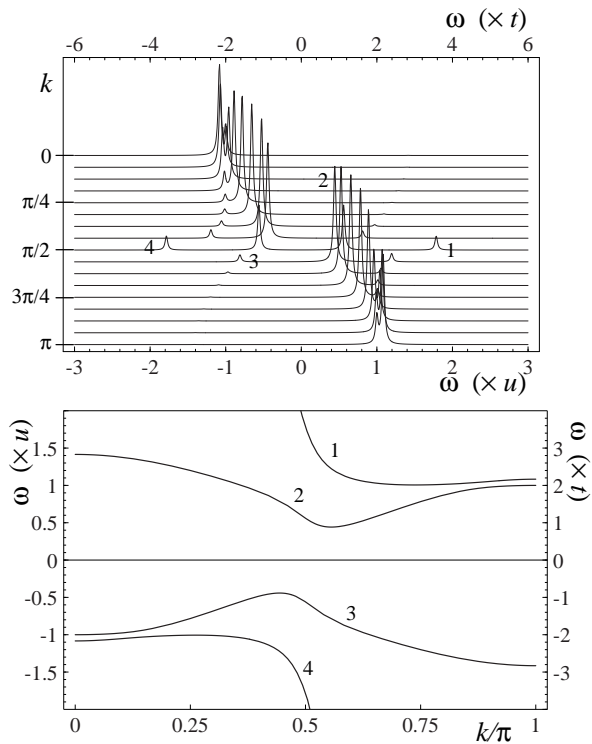}}
\caption{
(above)~Spectral function $A(k,\om)$ for point C ($t=0.5u,\>T=0.025u$ or
$U=4t,\>T=0.05t,\>\be=20/t$) with $\eta =0.02$.
(below)~Dispersion relation of the peaks.
}
\label{spec2}
\end{figure}

If we decrease T further from point B, we enter a regime that is beyond the
domain of validity of our approach. Indeed, in contradiction with the results
of Refs.~\cite{Voit96,Favand97,Kim96}, the spectral function becomes similar
to that of free particles, following a $-2t\cos k$ dispersion, except for a
gap at the Fermi energy for $k\simeq \pi/2$. We expect our expansion to be
valid if the $b_l(t)$'s in Eq.~(\ref{Jacob2}) are small compared to $\om$,
whose lowest-order value is $u$. This leads us to the conditions $t/u\lesssim
0.5$ and $3(t/u)^3\lesssim T/u$, fulfilled by the points under the dashed line
in Fig.~\ref{phases}. However, these conditions may be too stringent because
the $t\rightarrow\infty$ limit also happens to be correctly given by our
solution Eq.~(\ref{Jacob2}). Furthermore, a free particle dispersion relation,
with a gap opened at the Fermi level, is what is expected at large $t$ and
small $T$ for an itinerant antiferromagnet. Fig.~\ref{spec2} (point C)
illustrates this behaviour. The parameters have the same value as in the
Monte-Carlo (MC) calculations of Ref.~\onlinecite{Preuss94} ($U=4t$,
$\be=20/t$). The general distribution of the spectral weight, and the
dispersion relation of the peaks~\onlinecite{Preuss94} are well accounted for
by our solution. We believe that peak 1 contributes to the large uncertainty
(due mainly to the Maximum Entropy Method itself) on the maxima of $A(k,\om)$
reported in Fig. 2 of Ref.~\onlinecite{Preuss94} for $k$ near $0$ and $\pi$.
For other values of $k$, peak 1 could not be resolved in
Ref.~\onlinecite{Preuss94} because of its small weight and because of the
magnitude of the time slice, unlikely to detect high-energy features. Thus,
our results for point C appear correct. Our method definitely fails in the
shaded area of Fig.~\ref{phases}, where spin-charge separation occurs, but
outside this region our solution is reliable under the dashed line, and
uncontrolled (but not necessarily bad) above it.

The cuprate chain material $\rm SrCuO_2$ studied in Ref.~\cite{Kim96} happens
to fall in the shaded regime. Nevertheless, our results allow us to predict
that features dispersing on a scale $J$, like peak 3 in Fig.~\ref{spec1},
should appear at $\pi/2 \le k \le \pi$ upon raising T. Hints of this finite T
effect have already been seen in the ``question-mark'' features in Fig. 1 of
Ref.~\cite{Kim96}.

In summary, we presented a general method for constructing
strong-coupling expansions and applied it to the half-filled Hubbard
model. We showed how the Mott transition and AF correlations manifest
themselves in the single-particle properties. Finally, we gained
further insight into ongoing ARPES experiments on the propagation of
one hole in an AF correlated Mott insulator. Doping and two-particle
correlations are accessible within the same approach.

%

We thank C. Bourbonnais for numerous enlightening discussions.  We are
also grateful to  H. Touchette, L. Chen and S. Moukouri for sharing their
numerical results. This work was partially supported by NSERC (Canada), by
FCAR (Qu\'ebec),  by a scholarship from MESR (France) to S.P. and (for
A.-M.S.T.)  by the Canadian Institute for Advanced Research. 

%


\begin{references} 
%
\bibitem[*]{Stef} Also at the {\it Laboratoire de Physique des Solides, 
Universit\'e Paris-Sud, 
91405, Orsay, France}. 

\bibitem{Wells95} B. O. Wells {\it et al}, Phys. Rev. Lett. {\bf 74}, 
964 (1995). 
 
 
\bibitem{Kim96} C. Kim {\it et al}, Phys. Rev. Lett. {\bf 77}, 4054 (1996). 
 
\bibitem{Hubbard63} J. Hubbard, 
Proc. R. Soc. (London) Ser. A {\bf 276}, 238 (1963), 
A {\bf 277}, 237 (1964), 
A {\bf 281}, 401 (1964), 
A {\bf 285}, 542 (1965). 
 
\bibitem{Lieb68} E. H. Lieb, F. Y. Wu, Phys. Rev. Lett. {\bf 20}, 
1445 (1968). 
 
\bibitem{Preuss94} R. Preuss {\it et al}, Phys. Rev. Lett. {\bf 73}, 
732 (1994) and Ref.[14] therein. 
 
\bibitem{Preuss95} R. Preuss, W. Hanke, W. von der Linden, 
Phys. Rev. Lett. {\bf 75}, 1344 (1995). 

\bibitem{Bulut94} N. Bulut, D. J. Scalapino, S. R. White, Phys. Rev. B 
{\bf 50}, 7215 (1994). 
 
\bibitem{Favand97} J. Favand {\it et al}, Phys. Rev. B {\bf 55}, R4859 (1997). 
 
\bibitem{Emery79} V. J. Emery, {\it in} {\it Highly Conducting One-
Dimensional Solids}, 247, by J. T. Devreese {\it et al}, Plenum (1979).

\bibitem{Voit94} J. Voit, Rep. Prog. Phys. {\bf 57}, 977 (1994).

\bibitem{Voit96} J. Voit, cond-mat/9711064 (1997).

\bibitem{Solyom79} J. S\'olyom, Adv. Phys. {\bf 28}, 201 (1979).

\bibitem{Bourbon91} C. Bourbonnais, L. G. Caron, Int. J. Mod. Phys.
B {\bf 5}, 1033 (1991).

\bibitem{Frahm90} H. Frahm, V. E. Korepin, Phys. Rev. B {\bf 42},
10553 (1990); Phys. Rev. B {\bf 43},
5653 (1991). 

\bibitem{Kotliar92} W. Metzner, D. Vollhardt, Phys. Rev. Lett. {\bf 62}, 324 (1989).
M. J. Rozenberg, X. Y. Zhang, G. Kotliar, Phys. Rev. Lett. {\bf 69}, 1236 (1992).
A. Georges, W. Krauth, Phys. Rev. Lett. {\bf 69}, 1240 (1992).

\bibitem{Georges93} A. Georges, W. Krauth, Phys. Rev. B {\bf 48},
7167 (1993).

\bibitem{Pruschke93} Th. Pruschke, D. L. Cox, M. Jarell, Phys. Rev. B
{\bf 47}, 3553 (1993).

\bibitem{Georges96} A. Georges {\it et al}, Rev. Mod. Phys. {\bf 68}, 13 (1996).


\bibitem{Bartkowiak92} M. Bartkowiak, K. A. Chao, Phys. Rev. B {\bf 46}, 9228 (1992).

\bibitem{Metzner91} W. Metzner, Phys. Rev. B {\bf 43}, 8549 (1991).

\bibitem{Boies95} C. Bourbonnais, PhD thesis, Universit\'e de Sherbrooke,
(1985). D. Boies, C. Bourbonnais, A.-M. S. Tremblay,
Phys. Rev. Lett. {\bf 74}, 968 (1995).
S.K. Sarker, J. Phys. C: Solid State Phys. {\bf 21}, L667 (1988) (
the latter reference was pointed out to us by R. Fr\'esard).

\bibitem{Gilewicz78} J. Gilewicz, {\it Approximants de Pad\'e},
Lecture Notes in Mathematics {\bf 667}, Springer-Verlag (1978).

\bibitem{Moments} We call
$m_n({\bf k})=\int_{-\infty}^{+\infty}\om^n A({\bf k},\om)
d\om/2\pi,\>
n=0,1,2,...$ the moments of the spectral function.

\bibitem{precision} difference in $O(t^4)$ for $a_0,a_1,a_2,b_1,b_2$,
$O(t^2)$ for $a_3,a_4,b_3,b_4$, $O(1)$ for all other coefficients.

\bibitem{d-infty} This simple criterion yields $U_{\rm c}=1.86 t^\star$ 
(with $t^\star=2t\sqrt d$) in
infinite dimension. This critical value of interaction strength 
is too large because when $d\to\infty$ our criterion corresponds to
subbands that meet with an exponentially small density of states,
therefore not yet truly closing the gap.

\end{references}
\end{document}